# Role of ionic liquids in protein refolding: native/fibrillar versus treated lysozyme

Sara Mangialardo,*[a] Lorenzo Gontrani,[b] Ruggero Caminiti,[c] Paolo Postorino[d]

[a] Dpt of Physics, Università di Roma Sapienza,Italy.Fax: +3906443158 Tel: 06499123502. E-mail: saramangialardo@fastwebnet.it
[b] CNR-ISM Tor Vergata, Roma, Italy.E-mail:l.gontrani@caspur.it
[c] Dpt of Chemistry,, Università di Roma Sapienza, Italy.E-mail: r.caminiti@caspur.it
[d] Dpt of Physics, Università di Roma Sapienza, Italy. Fax: +3906443158 Tel: 06499123502. E-mail: paolo.postorino@roma1.infn.it

Several ionic liquids (ILs) are known to revert aggregation processes and improve the *in vitro* refolding of denatured proteins. In this paper the capacity of a particular class of ammonium based ILs to act as refolding enhancers was tested using lysozyme as a model protein. Raman spectra of ILs treated fibrillar lysozyme as well as lysozyme in its native and fibrillar conformations were collected and carefully analyzed to characterize the refolding extent under the effect of the IL interaction. Results obtained confirm and largely extend the earlier knowledge on this class of protic ILs and indicate Ethyl Ammonium Nitrate (EAN) as the most promising additive for protein refolding. The experiment provides also the demonstration of the high potentiality of Raman spectroscopy as a comprehensive diagnostic tool in this field.

## Introduction

Protein aggregation is one of the most important problems in the production and storage industrial processes and therefore among the causes of the major economic loss in biotechnology and pharmaceutical factories. Indeed in manufacturing commercial products, the goal is to obtain a stable and correct protein folding for allowing the full functionality.[1] The problems of protein aggregation and structural stability are not limited to the manufacturing processes. Protein misfolding diseases are a well-known class of ailments including Alzheimer, Parkinson and Huntington diseases. They all involve protein aggregation and share common features such as the presence of insoluble fibrous protein aggregation in a specific structural motif characterized by a cross-β sheet structure.[2] Key issues on aggregation are not yet fully addressed such as the detailed microscopic mechanism leading to aggregation, the structure of aggregates, how the environmental conditions can affect the rate and the amount of aggregation and how aggregation can be prevented and/or removed.

Additives may promote the stabilization of the native state of the protein accelerating the kinetics of the correct folding and removing/inhibiting the aggregation of denatured polypeptides and intermediates of the folding pathways.[3]

In recent years ionic liquids (ILs) have been used to stabilize the protein activity, to inhibit or reduce aggregation, and to improve the *in vitro* refolding of denatured proteins.[4,5] ILs have numerous attractive characteristics including their non-volatility, good solvating properties, thermal stability, and recyclability, that render these compounds "environmentally green".[6,7,8,9,10] One of the most important qualities of these solvents is the high tunability of their chemical structure. Indeed, they can be designed to have specific physical and chemical qualities by acting on either the alkyl chains (i.e. modifying the length, the presence of hydrophobic groups, etc.) or the anion (i.e. varying the degree of the charge delocalization, its hydrogen bonding ability, etc).[11] ILs having coordinating anions which are strong hydrogen bond acceptors (e.g. $Cl^-$, $NO_3^-$, $CH_3COO^-$ and $(MeO)_2PO_2^-$) can dissolve many compounds which are insoluble or sparingly soluble in water and in most of the organic solvents. Examples include cellulose[12] and several compounds having specific pharmacological activity.[13,14,15] As to the effects of the cation alkyl chain length, it has been proposed that the polarity of ILs decreases with increasing the alkyl chain length.[10] This can be an important chemical parameter as the polarity of ILs have an impact on the enzyme stability and selectivity. In particular the lengthening of the alkyl chain seems to have a positive effect on dissolving the aggregates and a negative effect on refolding the protein.[16,17,18,19]

In the last years the effects of ILs on several proteins have been extensively investigated by varying ILs composition (anion, cation, and alkyl chains) and denaturing conditions up to fibrillar aggregates.[20,21,22] Many interesting studies on the ILs-protein interaction have been carried out on lysozyme (Fig. 1), an enzyme often used as a model protein to study fibril formation *in vitro*.[23,24] As a matter of fact this protein can be easily converted to amyloid fibrils under high temperature and low pH environmental conditions and the fibrils formed share similar characteristics to ailment's amyloid.[25] Recently N. Byrne and C.A. Angell[26] have found that a protic IL (PIL), namely ethyl ammonium nitrate (EAN), has the property of partly recovering lysozyme functionality also after severe denaturing procedures leading to fibrils formation. This is a particularly interesting finding since it is rather uncommon to find an additive able to dissolve aggregates and contemporary to refold, at least partly, the protein. We want to stress that in ref. 26, as well as in most of the relevant literature, the enzymatic activity and the amount of protein aggregates have been exploited as a measure of the ILs ability as a refolding enhancers and aggregation inhibitor. A specific analysis of the structural rearrangements induced by the ILs interaction is therefore lacking in most cases.

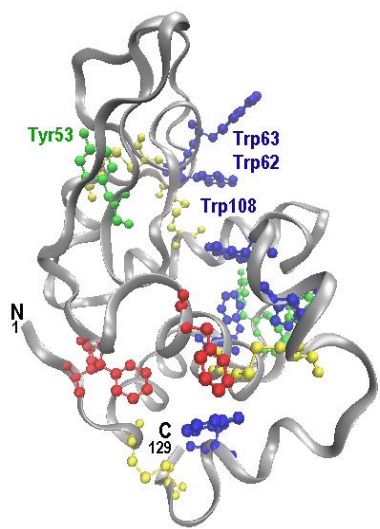

**Figure 1** - Hen Egg White Lysozyme (HEWL). N terminal and C terminal are labeled. The aromatic residues that were discussed in the text are highlighted: Phe (red), Tyr (green), Trp (blue). Among Cysteine residues (yellow) the disulphide bridges take place. Tyr53, Trp62, Trp63 and Trp108 are in the active site of the enzyme.

In the last decades Raman spectroscopy has become a common practice in protein structural investigations. This technique has many advantages: it is a non-destructive and non-invasive technique and it requires small amount of sample. Moreover, reliable assignments between specific spectral features and local protein structures have been established, and clear, sensitive spectroscopic markers of the secondary structure[27] and of the side-chain environments[28] have been identified. A careful data analysis of Raman spectra can also provide direct information on the protein tertiary structure (e.g. on disulphide bridges[29] and hydrogen bonds of the side chains[30]). The analysis of the amide bands is almost routinely exploited for the empirical-quantitative estimate of the protein secondary structure[31] with an accuracy comparable with that obtainable from circular dichroism experiments as witnessed by the good agreement usually found with the analysis of X-ray structural data.[32]

In the present paper we report on a careful Raman study of hen egg white lysozyme (HEWL) in both native and fibrillar conformations as well as on several PILs treated fibrillar HEWL. This study allowed us to get a full spectroscopic characterization of the protein in the two extreme conformations and, by comparison, to evaluate the refolding efficiency of EAN and of many other PILs of the same class of ammonium based PILs. Taking full advantage of the potentiality of the Raman spectroscopy as a structural diagnostic tool, we were able to confirm the capability of EAN as a refolding enhancer (see ref. 26) and, in addition, to achieve a deeper insight of the EAN induced structural modifications of lysozyme. The relevance of the alkyl chain length on the refolding efficiency of the different PILs investigated is finally briefly discussed.

## Experimental

### Materials

Hen egg white lysozyme powder was purchased from Fluka (62970) and used without further purification. HEWL fibrils were prepared by following a thermo-chemical protocol.[33] A quantity of 14 mg of HEWL were dissolved in 1 ml of distilled, purified water and the pH of the solution was adjusted to 2 by adding HCl. The temperature was then gradually increased with a rate of 10°C / 1h up to 72°C and the solution was kept at this temperature for 6 days.

The PILs used in the present work are based on the nitrate anion, $NO_3^-$.[34,35,36] Their typical formula is schematically depicted in Fig. 2, while their complete formulas are reported in Table 1. EAN and PAN were acquired from IoLiTec (Ionic Liquids Technologies), while BAN and MEOAN were synthesized in house.

Following ref. 26 weighted samples (1 mg) of HEWL fibrils were placed in vials and an aliquot (1 ml) of PILs were added. After 20 minutes at ambient conditions the solutions were centrifuged for 3 minutes, decanted and washed for 3 times. Finally samples were dried to be measured by Raman spectroscopy.

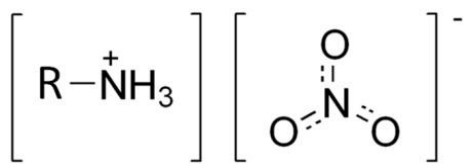

| Name | Formula |
|---|---|
| 2-Methyl Oxy Ethyl Ammonium Nitrate (MEOAN) | $[CH_3OCH_2CH_2N^+H_3][NO_3]^-$ |
| Ethyl Ammonium Nitrate (EAN) | $[CH_3CH_2N^+H_3][NO_3]^-$ |
| Propyl Ammonium Nitrate (PAN) | $[CH_3CH_2CH_2N^+H_3][NO_3]^-$ |
| Butyl Ammonium Nitrate (BAN) | $[CH_3CH_2CH_2CH_2N^+H_3][NO_3]^-$ |

**Figure 2** - General formula of the four PILs used in this work.

**Table 1** - Names, acronyms and complete chemical formulas of the four PILs used in this work.

### Methods

Raman measurements were carried out using a confocal micro-Raman spectrometer by Jobin-Yvon, equipped with several objectives, a 20 mW He–Ne laser (632.8 nm wavelength), and a 1800 lines mm$^{-1}$ grating. Raman spectra were collected in the back-scattering geometry and a notch filter was used to reject the elastic contribution, thus preventing also the collection of spectra close to the excitation line. Raman spectra were collected by means of a Peltier-cooled CCD (charge coupled device). Measurements were performed separately over four spectral ranges to cover the 200-3600 cm$^{-1}$ wavenumber region with a resolution better than 3 cm$^{-1}$. A large confocal diaphragm of 400 μm has been used in order to obtain a good Raman signal. The absolute wavenumber calibration for each spectral range was obtained by collecting the emission lines of a neon lamp. Further experimental details can be found in ref. [37].

Preliminary measurements on the protein powder placed onto a quartz slide were performed using available filters and objectives to find the best experimental conditions. Sample homogeneity and the absence of polarization effects were proved by repeating measurements on different sample regions. The typical acquisition time was 10 minutes for each frequency range. As a reference the Raman spectra of the PILs were collected in a quartz cuvette (1mm of optical path) after being de-hydrated in controlled atmosphere under nitrogen flux.

All the spectra were fitted using Levenberg-Marquardt minimization algorithm (LM algorithm) and Lorentzian-Gaussian pseudo-Voigt functions as peak profiles.[38,39,40,41]

## Results and Discussion

### Raman Spectra of Native and Fibrillar Lysozyme

Before discussing our spectroscopic data we want to recall that protein fibrillation pathway starts with the destabilization and partial unfolding of the native protein induced by high temperature and low pH environment. Partial unfolded proteins are converted into intermediates oligomers which subsequently are turned into protofibrils and finally into amyloids.[42] The sequence of these structural transitions can be monitored by Raman spectroscopy which allows to follow the protein tertiary structure by means of the skeletal bending and the C-C-N stretching frequencies of the peptide backbone, and by the S-S and the C-S stretching frequencies of the disulphide bonds. The spectra of native powder and fibrillar HEWL shown in Fig. 3 reveal indeed remarkable differences. The peaks ascribed to the S-S stretching vibrations between 500-550 cm$^{-1}$ (υ(S-S) in Fig. 3) are still present after the thermo-chemical process, even if their shapes are clearly modified. The correlation between the band frequencies and the conformers of the disulfide bonds has been well established through normal coordinate analysis and extensive experimental investigation.[43,44] As shown in Fig. 3, the four S-S bridges in native lysozyme give rise to three Raman bands at 507, 526 and 540 cm$^{-1}$ indicating that the intra-molecular S-S bonds in native lysozyme are in GGG, GGT and TGT conformations,[45] in agreement with the results of Van Wart et al.[29] After fibril formation, the intensities of υ(S-S) vibrations near 530 and 540 cm$^{-1}$ disappeared, clearly indicating a distortion of the dihedral angles with respect to the native structure.

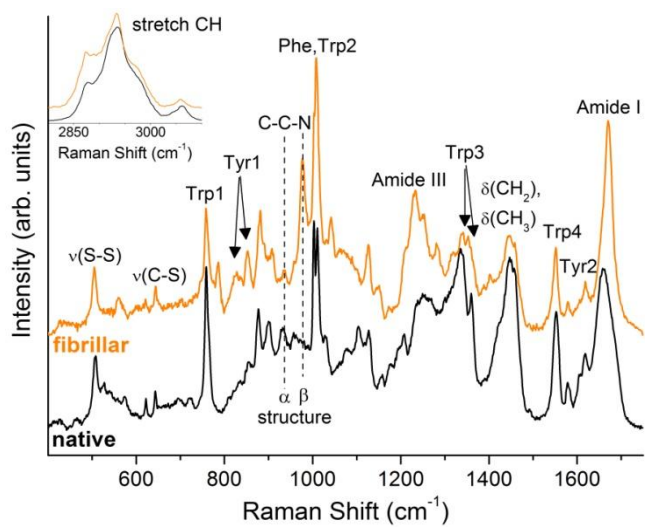

**Figure 3 -** Raman spectra of HEWL in native and fibrillar conformations. The region of CH stretching mode is shown in the inset. The spectra were normalized in the frequency region of the Phe and Trp2 peaks (990-1025 cm$^{-1}$).

On the other side, the C-C-N stretching peaks are centered around 930 cm$^{-1}$ when the protein structure is mainly α helix and at higher frequencies (around 960-980 cm$^{-1}$) when it is mainly in a β sheet conformation (see the vertical dashed lines in Fig. 3).[46,47] A close inspection of Fig. 3 clearly reveals that the band intensities ascribed to α helix and a β sheet conformations are markedly unbalanced towards the β structure on going from the native to the fibrillar protein.

An in-depth analysis of the Raman spectra allows to monitor the environment experienced by side chains and, consequently, to get information on the protein folding. This is the case of the analysis of the relative intensities of the peaks forming the Fermi doublets[48] arising from Tyr (830-850 cm$^{-1}$) and Trp residues (1340-1360 cm$^{-1}$) clearly detectable in the Raman spectra of both native and fibrillar HEWL (Tyr1 and Trp3 in Fig. 3, respectively).[30,49]

The peak intensity ratio $I_{850}/I_{830}$ of Tyr1 is a sensible marker of the hydrogen bonding of the Tyr phenoxyl group. If the ratio is around 0.3 the phenolic hydroxyl is the proton donor in a strong hydrogen bond, if $I_{850}/I_{830}$ is around 1.3 then the phenolic oxygen is both an acceptor and a donor of a weak hydrogen bond, and if the ratio is around 2.5 then the phenolic oxygen is the acceptor of a strong hydrogen bond.[50] A decrease of $I_{850}/I_{830}$ ratio reflect an increase buriedness, while a ratio around 1.3 is typical of a solvent exposed residues. The analysis of our data shows $I_{850}/I_{830} \cong 1.3$ for the native sample accordingly with a conformation where Tyr residues are exposed and able to participate in moderate or weak hydrogen bonding.[51] After the thermo-chemical treatment the ratio exceeds 2.0 suggesting the onset of strong hydrogen bonds and a conformation with Tyr more exposed to the solvent. This can be seen as a further spectroscopic signature of the unfolding process occurred after the thermo-chemical treatment.[52,53] It is worth to notice that the displacement of Tyr side chains in lysozyme during the unfolding of the tertiary structure can be strictly related to a change in the activity of the enzyme. Indeed, Tyr53 (see Fig. 1) is hydrogen bonded with the amino acid group of Asp66 and it is adjacent to the catalytic residue Asp52. A change of the Tyr53 position could thus affect the enzymatic active site of lysozyme.[54]

The peak intensity ratio $I_{1360}/I_{1340}$ of the Trp3 doublet is a marker of the hydrophobic/hydrophilic environment of the Trp indole ring.[55] If the relative intensity ratio $I_{1360}/I_{1340}$ is smaller than 1.0, the indole ring is considered to be in a hydrophilic environment (or exposed to aqueous medium) whereas if the ratio is greater than 1.0, it is considered to be in a hydrophobic environment (or in contact with aliphatic side chains). The analysis of the present measurements provides for the $I_{1360}/I_{1340}$ ratio the values of 0.4 for the native and 2.5 for the fibrillar sample. The Trp side chains thus pass from a hydrophilic to a hydrophobic environment on going from native to fibrillar state.

Changes in the Raman spectra in the regions of Trp residues are of great importance because 3 of the 6 Trp residues (Trp62, Trp63 and Trp108, see Fig. 1) are in the active site of the enzyme,[56] thus changes in their environment should play an important role in the enzymatic activity. Other important information on lysozyme environment and structure can be obtained analyzing the other bands associated with the Trp residues at 1550 and 1011 cm$^{-1}$ (Trp4 and Trp2 in Fig. 3). The latter shows a frequency shift from 1011 cm$^{-1}$ (native) to 1008 cm$^{-1}$ (fibrillar)[57] although, in this case, the conspicuous redistribution of the spectral intensities between the Trp2 and the close Phe peaks (~ 1003 cm$^{-1}$) is the most remarkable effect. This finding can be ascribed to a decrease of the intensity of the Phe peak rather than to an increase of the Trp2 peak thus suggesting a larger exposure of Phe residue to the solvent after the thermal treatment.[58]

HEWL is a globular protein consisting of approximately 40-45% of α helix and around 20% of β sheets in its native conformation.[59] Since it is well known that different secondary structures give rise to different components in the Raman amide bands,[60] a standard analysis was carried out to obtain quantitative information on the secondary structures of HEWL. The results of the LM fitting procedure of the amide I band for the native structure are in good agreement with the literature values[61] and are reported together with those obtained for the fibrillar sample in Table 2 (the spectral deconvolution of the amide band of the untreated fibrillar sample is shown in the following section in Fig. 5A). For the secondary structure characterization, since we are mainly interested in the α helix and β sheet contributions, we categorize the rest of the protein conformations as unordered structures. The fibrillation process decreases the α helix and increases the β components. It has been previously reported that the random component of the secondary structure plays a relevant role in the fibrillation process, as in the first stage of the fibrillation the native structure falls in a disordered structure and then develops into an organized β inter-chain structure.[62] It is important to recall that the peak at around 1670 cm$^{-1}$ is the marker of ordered β sheet structures in the aggregated sample, and that this feature together with the narrowing of the amide I band are the most clear markers of fibril formation in the Raman spectra.[63] Shifts at lower frequencies can be observed in the amide III region coherently with the above mentioned changes in the secondary structure from α helix (around 1300 cm$^{-1}$) toward ordered β sheet (around 1230 cm$^{-1}$) (see Fig. 3).[39,64]

|  | Native HEWL | Fibrillar HEWL |  |
|---|---|---|---|
| Unordered (random + turns) | 1644, 1693 | 1648, 1689 | Wavenumber (cm$^{-1}$) |
|  | 32 | 19 | Area (%) |
| α helix | 1660 | 1659 | Wavenumber (cm$^{-1}$) |
|  | 45 | 19 | Area (%) |
| β sheet inter-chain | - | 1670 | Wavenumber (cm$^{-1}$) |
|  | 0 | 39 | Area (%) |
| β sheet intra-chain | 1679 | 1633, 1679 | Wavenumber (cm$^{-1}$) |
|  | 23 | 23 | Area (%) |

**Table 2** - Results from the fitting procedure of the Amide I band of the HEWL sample in both native and fibrillar states.

## Raman Spectra of HEWL treated with PILs

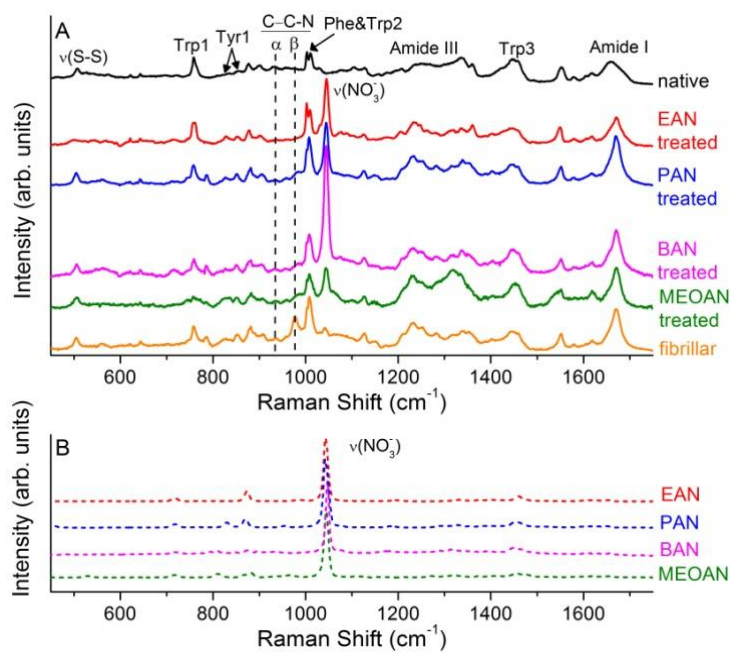

**Figure 4 -** A) Raman spectra of the HEWL in native and fibrillar conformation, and of the fibrillar sample treated with: EAN, PAN, BAN and MEOAN. B) Raman spectra of EAN, PAN, BAN and MEOAN.

Raman spectra of fibrillar lysozyme after the treatment with PILs are shown in Fig. 4A together with the spectra of the protein in the native and in the pristine fibrillar conformation. The spectra of the four pure PILs are shown Fig. 4B where it can be noticed that the symmetric stretching mode of $NO_3^-$ is largely the most prominent spectral feature.[65] Despite three successive centrifugations and re-dilutions in water, this peak is still present in the Raman spectra of the treated samples (see Fig. 4A). Nevertheless by comparing the spectra in Fig. 4A and 4B it is rather clear that the presence of a small amount of PIL in the treated samples does not significantly affect the Raman signal arising from lysozyme at least away from the frequency of $NO_3^-$ stretching peak. In particular the Raman contribution originating from residual PIL can be safely neglected over the frequency regions where the most relevant lysozyme spectral markers are found. By comparing the spectra of the treated samples in Fig. 4A with those collected from the protein in the fibrillar and the native conformations we can highlight and discuss the effects of the PILs different treatement.

From a qualitative point of view the Raman spectra of PAN and BAN treated samples appear very similar to the one collected from pristine fibrillar HEWL. In particular the υ(S-S) vibration of the disulphide bonds is lacking of the highest frequency components, indicating the same disposition of the dihedral angle than the fibrillar sample, distorted with respect to the native structure. Similar conclusions can be drawn looking at the intensity ratio found for the Tyr1 and Trp3 doublets. The values found for PAN and BAN treated samples show indeed that Tyr experienced a strong hydrogen bonds ($I_{850}/I_{830}$ = 1.9, and 1.8) and that Trp are still in a hydrophobic environment ($I_{1360}/I_{1340}$ = 1.9 and 1.5) as well as in the pristine fibrillar sample ($I_{850}/I_{830}$ =2.1 and $I_{1360}/I_{1340}$ =2.5). Some differences with the pristine fibrillar sample can be found in the C-C-N stretching peak region where a broadening and an intensity reduction of the spectral structure around 977 cm$^{-1}$ (β anti parallel configuration) is observed in the two treated samples.[46] These findings reflect slight changes in the tertiary structure and reveal an increased disorder in the β structures of the treated samples (see also the quantitative analysis of the amide I band reported in the following).

A similar peak broadening and intensity decrease can be observed in the spectrum of the MEOAN treated sample, albeit, in this case other spectral modifications are also evident from the comparison with the spectrum of the pristine fibrillar lysozyme. In particular MEOAN modifies the lysozyme tertiary structure mainly on Phe, Tyr and Trp residues. The interaction between MEOAN and the aromatic residues is indeed well evidenced in the Raman spectra by the modifications of the frequencies, shapes, and intensities of the

| | Fibrillar HEWL | Fibrillar HEWL treated with MEOAN | Fibrillar HEWL treated with BAN | Fibrillar HEWL treated with PAN | Fibrillar HEWL treated with EAN | Native HEWL | |
|---|---|---|---|---|---|---|---|
| Unordered (random + turns) | 19 | 11 | 20 | 20 | 27 | 32 | Area (%) |
| α helix | 19 | 16 | 14 | 14 | 31 | 45 | Area (%) |
| β sheet inter-chain | 39 | 39 | 35 | 35 | 30 | 0 | Area (%) |
| β sheet intra-chain | 23 | 34 | 33 | 31 | 12 | 23 | Area (%) |

**Table 3 -** Result of the fitting procedure of the amide I band of the HEWL sample in both native and fibrillar states.

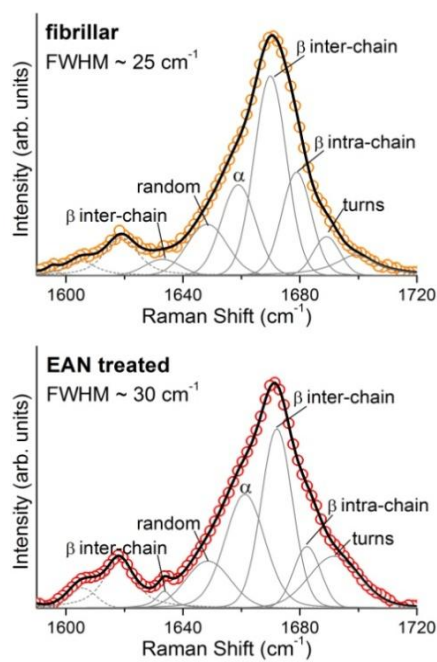

**Figure 5** - Raman spectra of the fibril HEWL (orange circles) and the fibril HEWL treated with EAN (red circles) fitted through the LMA (black line in both panels). In solid grey the different conformations of the amide I band are shown and in dashed grey are shown the Tyr2 and Phe contributions.

Trp peaks at 757 cm$^{-1}$ (Trp1), 1340/1360 cm$^{-1}$ (Trp3), 1550 cm$^{-1}$ (Trp4) and Tyr peaks at 830/850 cm$^{-1}$ (Tyr1), 1610 cm$^{-1}$ (Tyr2). The frequency values here reported refer to the native conformation. To explain these changes in the Raman spectra we hypothesize a cation-π interaction between the NH$^+$ cation of MEOAN and the electrostatic negative charge of the aromatic ring of Phe, Tyr and Trp residues.[66,67] MEOAN is the only PILs among those here studied that exhibits such a behavior, we can thus address this peculiar interaction to the presence of an ether group in the alkyl chain which makes MEOAN more polar than the others. On the other hand this type of interaction is quite common in proteins[ref] and we have observed similar spectral modifications in a sample of fibrillar insulin treated with MEOAN (data not shown). Moreover the Raman spectrum of the treated MEOAN lysozyme shows changes with respect to that of the pristine fibrillar sample also in the peaks involved in the NH modes (see around 1317 and 1540 cm$^{-1}$), reinforcing the hypothesis of a cation-π interaction.

Among those presently investigated, EAN is the ionic liquid showing the most remarkable differences between the spectra collected from the EAN treated and the pristine fibrillar lysozyme. These changes are mostly compatible with a PIL induced refolding process. Looking at Fig. 4A, it can be noticed that the C-C-N stretching mode, related to the β anti parallel structure (around 977 cm$^{-1}$) totally vanishes in the spectrum of the EAN treated sample closely resembling the spectrum of the native protein. Indications of the ongoing refolding process can be also found looking at the Phe-Trp2 doublet just above 1000 cm$^{-1}$. As it was already observed before, the fibrillation process exposed even more the Phe residues to the solvent and this leads to a decrease in the intensities of the Phe peaks in the Raman spectrum (see Fig. 4A).[33] Further indication can be obtained by the analysis of the Raman spectra in the S-S stretching region and from the Tyr Fermi doublet. In both cases the EAN treated sample resembles the native lysozyme. $\upsilon$(S-S) reveal a restoration of the pristine dihedral angles, while the value of the Tyr1 $I_{850}/I_{830}$ ratio is 0.9, close to the native one and displaying a character in the hydrogen bond that is both donor and acceptor. Finally the value obtained by the analysis of the ratio of the Trp3 Fermi doublet ($I_{1360}/I_{1340}$=0.9) shows a partial restore of the hydrophilic environment around the Trp residues, resembling what is for the native lysozyme. We also would like to notice that the peak around 786 cm$^{-1}$ (see Fig. 4A), albeit not clearly assigned, can be found in all the spectra but the native and the EAN treated ones.

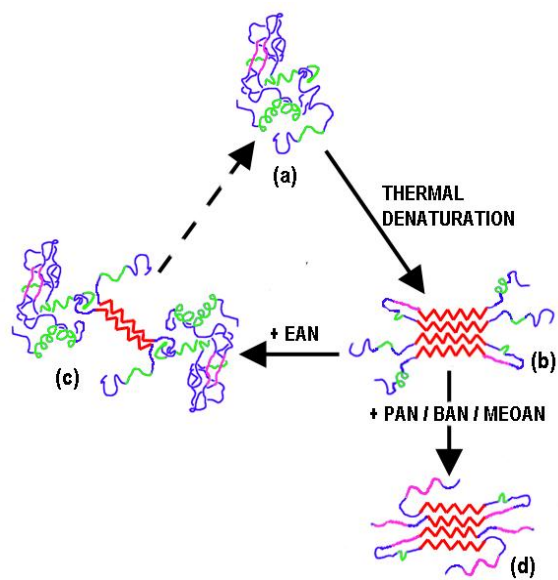

**Figure 6 -** Schematics of HEWL conformations: (a) native, (b) fibrillar, (c) partially refolded after EAN treatment, and (d) average configuration after PAN, BAN, and MEOAN treatments. α helices segments are in green, β intra-chain in magenta, β inter-chain in red, and unordered segments in blue.

Looking at Fig. 4A we notice that the amide I band is actually unaffected by any PILs treatments except for the EAN which causes a detectable broadening the band: the FWHM (full width half maximum) goes from 25 cm$^{-1}$ to 30 cm$^{-1}$ after the EAN treatment of the fibrillar sample. As mentioned above, more quantitative information about the refolding process can be obtained by a careful shape analysis of the amide band. A standard band fitting procedure was indeed carried out for all the investigated samples. In Fig. 5 the best fit curves obtained using the LM algorithm for the EAN treated and the pristine fibrillar samples are compared with the experimental data. The different components used in the fitting procedure associated to different secondary structures of the HEWL conformation are also shown. In Table 3 the percentages for the secondary structures considered are reported. Also in this case the analysis shows the peculiarity of the EAN treatment which apparently induces modification of the secondary structure of fibrillar HEWL different from those induced by the other PILs. In particular, going from the fibrillar to the EAN treated sample the percentage of the α helix grows (from 19% to 31%) and that of β sheet inter-chain diminishes (from 39% to 30 %). It is important to stress once more that the concomitant diminishing of the β sheet inter-chain structure and the broadening of the amide I band (see Fig. 5) in the EAN treated sample are clear signs of the ongoing refolding process. The behavior of fibrillar lysozyme in EAN is in agreement with the recovery of the enzymatic activity observed by Byrne et al. (ref. 26). Looking at Table 3 it is also important to notice that, albeit weak, a systematic decrease of the unordered components simultaneous to the increase of the β sheet intra-chain trend can be observed going from PAN to BAN to MEOAN, i.e. actually on increasing the alkyl chain.

## Conclusions

We reported on a complete Raman investigation of lysozyme in the native and fibrillar conformations. In addition to the standard shape analysis of the amide I band, the backbone deformation and the Fermi doublets of Tyr and Trp were deeply analyzed thus obtaining a comprehensive spectral characterization of the native and fibrillar lysozyme. This preliminary investigation provided us of the fundamental basis for an analysis of the refolding effect of several PILs on fibrillar lysozyme. We indeed carried out a systematic Raman investigation of four PILs treated fibrillar lysozyme samples and through the comparison with the native and the untreated fibrillar sample we were able to evaluate the refolding efficiency of the solvents.

The comparative spectral analysis shows that PAN and BAN (long chain PILs) only partially affect the tertiary structure of the fibrillar protein leaving the secondary structure actually untouched from the pristine fibrillar conformation. No evidences of a real refolding process have been found also in the fibrillar sample treated with MEOAN (a long chain PIL with an ether group). In this case, likely due to the presence of the ether group, we observed a clear indication of a different PIL-protein interaction mainly affecting the aromatic residues. Our results show that only the treatment with EAN (the shortest chain PIL) induces a significant refolding of fibrillar lysozyme.

Schematics of the protein conformations induced by the thermo-chemical and by the subsequent PIL treatments are shown in Fig. 6.We notice that the fibrillar conformation (Fig.6b) is mainly characterized by a large extent of β aggregates (β inter-chain, red segment) which, according with the results of the quantitative analysis of the secondary structure (amide band), is significantly reduced by the EAN treatment (Fig. 6c). This progress towards the native conformation (Fig.6a) induced by the EAN treatment is accompanied by the refolding of several protein structures (blue and green segments in Fig.6c) consistently with the indications obtained from several spectroscopic markers related to specific residues. In particular Tyr and Phe during the fibrillation process are more exposed to the solvent, while Trp changes its state from an exposed state in the native conformation to a buried state in the fibrillar. The EAN treatment report all the residues to their pristine conformations: Tyr and Phe are found to be less exposed to the solvent, while Trp residues goes back to the exposed state typical of a native conformation. These changes are particularly relevant since Tyr53, Trp62, Trp63 and Trp108 are in the lysozyme active site and have an important role in its enzymatic activity.[54,56] This result is fair good agreement with the recovery of the lysozyme functionality reported in ref. 26.

Treatments with long chain PILs (PAN, BAN, and MEOAN) do not drive fibrillar lysozyme toward the native conformation but, on the contrary, they increase the percentage of β intra-chain (magenta segments) with respect to both fibrillar and native conformations. Generally speaking the interaction with long chain PILs leads to a protein conformation markedly β (over 70% on average) far away from the mainly α helix structure of the native protein (45% α and 23% β) but also from the fibrillar (~ 60% β).We want to stress that the analysis of the spectroscopic markers related to Trp, Tyr and Phe residues shows that they are still in the fibrillar conformation thus preventing any recovery of the enzymatic activity.

The possibility offered by the Raman spectroscopy to gain a deeper insight into the structural modifications induced by refolding additives opens the way to a systematic use of Raman spectroscopy as a diagnostic tool in refolding studies, especially in determining the efficiency of IL-based additives for protein refolding.

On concluding, in the most recent literature it is widely accepted that the lengthening of the alkyl chain favors aggregate dissolution and inhibits protein refolding. Indeed, our results show a significant dependence of the PIL refolding enhancer properties from the cation structure, but the long alkyl chain compounds weakly affect the ordered β aggregates, peculiar of the fibrils conformation, thus preventing any conversion of the β sheets into α helices. On the contrary the shortest alkyl chain compound (EAN) is able to induce both the fibril melting and the refolding of the secondary structure. The information obtained by our analysis surely help in understanding the microscopic mechanism at the origin of the lysozyme functionality recovery.

**Acknowledgment**

L.G. acknowledges support from FIRB "Futuro in Ricerca" research project RBFR086BOQ_001, "Structure and dynamics of ionic liquids".